# Effect of randomly occurring Stone-Wales defects on mechanical properties of carbon nanotubes using atomistic simulation

Qiang Lu and Baidurya Bhattacharya[1]

*Department of Civil and Environmental Engineering, University of Delaware, Newark, DE 19716, USA*

**Abstract**

The remarkable mechanical properties of carbon nanotubes (CNTs) have generated a lot of interest in recent years. While CNTs are found to have ultra-high stiffness and strength, an enormous scatter is also observed in available laboratory results. This randomness is partly due to the presence of nanoscale defects, heterogeneities etc., and this paper studies the effects of randomly distributed Stone-Wales (SW or 5-7-7-5) defects on the mechanical properties of single-walled nanotubes (SWNTs) using the technique of atomistic simulation (AS). A Matern hard-core random field applied on a finite cylindrical surface is used to describe the spatial distribution of the Stone-Wales defects. We simulate a set of displacement controlled tensile loading up to fracture of SWNTs with (6,6) armchair and (10, 0) zigzag configurations and aspect ratio around 6. A modified Morse potential is adopted to model the interatomic forces. We found that fracture invariably initiates from a defect if one is present; for a defect-free tube the crack initiates at quite random locations. The force-displacement curve typically behaves almost linearly up to about half way, although there is no obvious yield point. Three mechanical properties – stiffness, ultimate strength and ultimate strain – are calculated from the simulated force and displacement time histories. The randomness in mechanical behavior resulting only from initial velocity distribution was found to be insignificant at room temperature. The mean values of stiffness, ultimate strength and ultimate strain of the tube decrease as the average number of defects increases although the coefficients of variation do not show such monotonic trend. The introduction of an additional defect has the most pronounced effect on the randomness in mechanical properties when the tube is originally defect free. We also found that, for a given mean number of defects in the tube, the zigzag configuration has less strength and less ultimate strain on the average, but more uncertainty in its stiffness and ultimate strain, compared with the armchair tube.

---

[1] Corresponding author. Email: bhattacharya@ce.udel.edu; fax 302 831 3640

# 1. Introduction

*1.1 Randomness in the mechanical properties of CNTs*

The report by Iijima [1] of the existence of helical carbon microtubules has triggered widespread interest of this special structure of carbon, now called carbon nanotubes (CNTs). Later, Ebbesen and Ajayan [2] reported the bulk synthesis of CNTs using a variant of the standard arc-discharge technique for fullerene synthesis under a helium atmosphere. Hamada et al. [3] and Saito et al. [4, 5] calculated the electronic properties of individual nanotubes. The Young's modulus was measured by Treacy et al. [6], Wong et al [7] and Falvo et al. [8]. An interesting statistics by Terrones [9] has shown an exponential growth of the number of publications on CNTs: from less than 50 in 1993 to more than 1500 in 2001. The study of carbon nanotube (CNT) has been motivated largely due to its extraordinary electronic and mechanical properties. CNT is found to be among the most robust materials: it has high elastic modulus (order of 1 TPa), high strength (up to 150 GPa), good ductility (up to 15% max strain), flexibility to bending and buckling and robustness under high pressure [10-13]. The combination of these properties makes the carbon nanotube a potentially very useful material and CNTs are now used as fibers in composites, scanning probe tips, field emission sources, electronic actuators, sensors, Li ion and hydrogen storage and other electronic devices. Also, CNTs can be coated or doped to alter their properties for further applications [9].

A survey of recent studies on the elastic modulus and strength of single-walled and multi-walled carbon nanotube (SWNT and MWNT) is listed in Table 1 and Table 2. The collected data clearly show a few features: (i) the Young's modulus of SWNT is found to range from 0.31~1.25TPa, the Young's modulus of MWNT ranges from 0.1 to 1.6 TPa; (ii) the Young's modulus may depend on chiralities; (iii) experiments roughly show a trend that Young's modulus drops quickly as the diameter of tubes increases; (iv) the strength varies a lot from about 5GPa to 150Gpa; (v) last but not the least important, these mechanical properties (elastic modulus, strength and failure strain) reported from experiments and analysis show *significant variation*. Such variation has also been noticed by a few recent studies [14, 15]. As detailed in the next section, some of the factors that give rise to randomness in *mechanical* properties affect CNT *electronic* properties as well. An analytical understanding of these variations, their sources and how they can be controlled is essential before CNTs and CNT-based products can be considered for widespread use across industries.

The huge scatter in these properties are possibly due to the variation in tube sizes (diameter and length), chiralities, random defects distribution, local fluctuations in energies, variations in the chemical environment, errors in measurements and even different ways in defining the cross-sectional area. Effects of sizes and chiralities of nanotubes on their mechanical properties have been studied [7, 16, 17], however, their roles are still not very clear. Table 1 and 2 also show that the elastic modulus measured from tubes with similar diameters varies quite significantly. Although experimental measurement techniques vary in sophistication, it is still doubtful that measurement errors are fully responsible for the huge variation. The lack of consensus in defining the cross-sectional area [18, 19] can be rectified with a simple multiplicative factor and is not likely to be a major source of the randomness. Therefore, a substantial part of the random effect is believed to occur from random defects and/or local energy fluctuations. In this study, we focus on the role of randomly occurring defects, particularly Stone-Wales defects, in the mechanical properties and fracture of CNTs.

**Table 1 A survey of mechanical properties of SWNTs**

| Ultimate strength (GPa) | Young's modulus (TPa) | Diameter (nm) | Length (nm) | Source, Method/Comment |
|---|---|---|---|---|
| NA | 1.25 -0.35 / +0.45 | 1.0 ~ 1.5 | 23.4 ~ 36.8 | [20] Thermal vibration experiment |
| NA | 0.97 ~ 1.20 | 0.4 ~ 3.4 | 10 | [21] AS using EAM potential. Modulus increases significantly with decreasing diameter and increase slightly with decreasing helicity. |
| NA | 0.8 ~ 1.22 | 0.8 ~ 2.0 | NA | [18] Ab initio calculation. Modulus depends on chiralities. |
| NA | 0.50 ~ 0.82 | 0.6 ~ 1.4 | Infinite[a] | [19] Ab initio calculation. Very little dependence on diameters and chiralities. |
| ≥45±7 | NA | 1.1 ~ 1.4 | >4000 | [22] AFM bending test on SWNT ropes. Maximum strain is measured as (5.8±0.9)%, strength is calculated by assuming E is 1.25 TPa |
| NA | 0.97 | 0.68~27 | NA | [23] Empirical force constant model |
| NA | 0.4~0.8 | 2.4~3.3 | NA | [24] AS using Brenner's potential, tensile loading |
| NA | 0.50 | 0.4~2.2 | Infinite[a] | [25] AS using Brenner's potential, tensile loading |
| NA | 0.98 | 1.3 | 14 | [26] Tight binding |
| NA | 1.2 | 3.1±0.2 | NA | [27] 3-point bending |
| NA | 5.0 | 0.68 | NA | [28] Tight binding calculation, wall thickness 0.7Å. |
| NA | 4.70 | NA | NA | [29] Local density approximation model, wall thickness 0.75Å. |
| 65~93 | NA | 1.6 | NA | [30] AS using modified Morse model, tensile loading |
| 62.9 | 0.83~3.02 | 0.68 | NA | [31] Atomistic simulation and ab initio calculation, w/o defects |
| 40~50 | 1.0 | 0.4 ~ 2.2 | 6 ~ 1000 | [32] Theoretical analysis based on AS. Modulus, yield strain and strength depend on loading rate. Yield strain depends on tube length. |
| 4.92 | 0.311 | 1.36 | NA | [33] Tight-binding simulation, maximum strain 22%. Poisson ratio 0.287. Modulus slightly depends on diameters. |

[a]Simulating infinite long tube with periodic boundary condition

**Table 2 A survey of mechanical properties of MWNTs**

| Ultimate strength (GPa) | Young's modulus (TPa) | Outer diameter (nm) | Length of tube (nm) | Method/Comment |
|---|---|---|---|---|
| 14.2 ± 8.0 | 1.28± 0.59 | 26-76 | 8.0 | [7] Buckling stress is measured, which should be smaller than tensile strength. No dependence of Young's modulus on tube diameter. |
| 100-150 | NA | 21 | Buckling: 18.3~68 Bending: 800 | [8] Both bending and buckling tests are conducted. The strength is calculated from a maximum strain of 16% from bending. |
| 55 | NA | 8.4-16.6 | 46.2~436 | [34] Critical buckling stress is measured. |
| 135-147 | NA | 19.6-56.2 | Buckling length: 47~223.5 | [35] Critical buckling stress is measured. |
| 11-63 | 0.27-0.95 | 13-33 | 6900~ 11000 | [36] Strength measured through direct stretching. Failure happens when the outermost layer is broken. No apparent dependence of tensile strength on the outer shell diameter. 12% strain at failure is measured. |
| 150 | 0.91±20% | <10 | About 500 | [37] Strength measured through direct stretching. |
| NA | 0.6-1.1 | 10~20 | NA | [6] Elastic modulus is measured through thermo vibration method, higher modulus is found for thinner tubes. |
| NA | 0.1-1.6 | 12~30 | 1500~6250 | [16] Elastic modulus is measured through resonant vibration, and found to decrease sharply with increasing diameter. |
| NA | <0.12 | 30~250 | About 2000 | [38] Elastic modulus is measured through resonant vibration, and found to increase rapidly as diameter decreases. |
| NA | 0.67-1.1 | 4.8~16.0 ±5% | 240~420 | [39] Elastic modulus is measured through beam deflection, and found no significant dependence on diameter. |

*1.2 Effects of defects on the mechanical properties CNTs*

Defects such as vacancies, metastable atoms, pentagons, heptagons, Stone-Wales (SW or 5-7-7-5) defects, heterogeneous atoms, discontinuities of walls, distortion in the packing configuration of CNT bundles, etc. are widely observed in CNTs [40-44]. Such defects can be the result of the manufacturing process itself: according to an STM observation of the SWNTs structure, about 10% of the samples were found to exhibit stable defect features under extended scanning [45]. Defects can also be introduced by mechanical loading [46, 47] and electron irradiation [40]. It is reasonable to believe that these defects are randomly distributed in CNTs.

Studies have shown that defects have significant influence on the formation as well as on the electronic and mechanical properties of CNTs [11, 43, 46]. It has also been suggested that oxygen sensitivity may be an effective metric of defect concentration in carbon nanotubes [48]. Pentagon and heptagon defects are believed to play key roles in the formation and deformation of CNTs. For example, with the help of pentagon and heptagons, one can build special structures based on the original hexagon lattice, such as capping, intramolecular junctions, variation in diameter or chirality, welding, coalescence, x-junction etc. [43].

The Stone-Wales (SW) defect, which is the focus of this paper, is composed two pentagon-heptagon pairs, and can be formed by rotating a $sp^2$ bond by 90 degrees (SW rotation). SW defects are stable and commonly present in carbon nanotubes [49], and are believed to play important roles in the mechanical [15, 50], electronic [51], chemical [52], and other properties of carbon nanotubes. For example, Chandra et al. [15] found that the SW defect significantly reduced the Elastic modulus of single-walled nanotubes. Mielke et al. [53] compared the role of various defects (vacancies, holes and SW defects) in fracture of carbon nanotubes, and found that various one- and two-atom vacancies can reduce the failure stresses by 14~26%. The SW defects were also found to reduce the strength and failure strain, although their influence was less significant than vacancies and holes.

It has been found that SWNTs, under certain conditions, respond to the mechanical stimuli via the spontaneous formation of SW defect beyond a certain value of applied strain around 5%~6% [47]. Atomistic simulations showed that these SW defects are formed when bond rotation in a graphitic network transforms four hexagons into two pentagons and two heptagons which is accompanied by elongation of the tube structure along the axis connecting the pentagons, and shrinking along the perpendicular direction. More interestingly, the SW defect can introduce successive SW rotations of different C-C bonds, which lead to gradual increase of tube length and shrinkage of tube diameter, resembling the necking phenomenon in tensile tests at macro scale. This process also gradually changes in chirality of the CNT, from armchair to zigzag direction. This whole response is plastic, with necking and growth of a "line defect", resembling the dislocation nucleation and moving in plastic deformation of crystal in many ways. Yakobson [54] thus applied dislocation theory and compared the brittle and ductile failure path after the nucleation of the SW defect.

The formation of SW defects due to mechanical strains has also been reported by other groups of researchers. In their atomistic simulation study, Liew et al. [50] showed that SW defects formed at 20~25% tensile strain for single-walled and multi-walled nanotubes with chirality ranging from (5,5) to (20,20). The formation of SW defects explained the plastic behavior of stress-strain curve. They also predicted failure strains of those tubes to be about 25.6%. A hybrid continuum/atomistic study by Jiang et al. [55] reported the nucleation of SW defects both under tension and torsion. The reported SW transformation critical tensile strain is 4.95%, and critical shear strain is 12%. The activation energy and formation energy of the SW defects formation are also studied and related to the strength of the nanotube [56-58]. The nucleation of SW defects was found to depend on the tube chiralities, diameters and external conditions such as temperature.

It is important to mentioned here that the above studies [47, 50, 54, 55] are all based on the bond order potential model. Two recent studies critical of the bond order model must also be mentioned in this context. Dumitrica et al. [59] argued that the energy barrier for SW defects formation at room temperature is high enough to inhibit stress-induced SW defects formation, and showed the direct bond-breaking is a more likely failure mechanism for defect-free nanotubes. A quantum mechanical study [60] from the same group pointed out that the results in [47] and related works were not reliable since the potential model they used (i.e., the Bond Order model) was not capable of correctly describing breaking of the bond connecting the two pentagons in the SW pair. They [60] further found that pre-existing SW defects caused successive bond breakings instead of bond rotations as reported by [47,

50, 54, 55]. Nevertheless, irrespective of how SW defects are formed and grow, it is clear that they can have significant effects of the mechanical properties of carbon nanotubes.

In conclusion, it is clear that (i) defects are commonly present in CNTs, (ii) these defects may have significant effects on the mechanical and other properties of CNTs, and as summarized in Table 1 and Table 2, (iii) a wide scatter has been observed in the mechanical properties of carbon nanotubes. Yet, surprisingly little work has been directed in the available literature toward studying the randomness in these defects and the influence of such randomness on CNT mechanical properties in a systematic and probabilistic way. Here in this paper, we try to build toward this missing link by selecting the effect of randomly occurring Stone-Wales defect in SWNTs as a potentially significant case study. We make reasonable assumptions on the random nature of the SW defects and, through the technique of atomistic simulation, quantify the effect of such randomness on (i) the fracture process of SWNTs at the atomic scale, and (ii) three representative mechanical properties, namely, the elastic modulus, ultimate strength and ultimate strain of SWNTs.

## 2. Modeling of random Stone-Wales defects

An ideal modeling of material defects would require a real 3-D random field model. However, since each SW defect on a SWNT is a local rearrangement of carbon atoms on the tube wall, we can treat the SW defects as disks with a fixed area that are randomly distributed on the 2D surface. This way, it is sufficient to consider two characteristics of the defects distribution: (i) The location of the points relative to the study area, and (ii) the location of points relative to each other.

To our knowledge, there are few published works to date that study the effects of random defects, especially of the Stone-Wales kind, on the mechanical properties of CNTs. The study by Saether [44] investigated the transverse mechanical properties of CNT bundles subject to random distortions in their packing configuration. This distortion, quantified by a vector describing the transverse displacement of the CNTs, may be caused by packing faults or inclusions. The magnitude and direction of the vector were both uniform random variables. The transverse moduli of CNT bundles were found to be highly sensitive to small distortions in the packing configuration. In another instance, Belavin et al [61] studied the effect of random atomic vacancies on the electronic properties of CNTs. Each of the $N_a$ atoms in the "unit cell" was assigned (using an unspecified numbering scheme) a unique id in the range $[1,N_a]$. A first estimate of the number of atoms to be removed was made by generating a uniform deviate, $n_d$, in the interval $[1,N_a]$. It is not apparent how the $n_d$ atoms to be removed were identified. However, the atoms in this set were removed from the unit cell provided the cell was left with atoms with coordination numbers of 2 or 3. That is, (i) the removal of an atom did not give rise to an additional dangling bond, or, (ii) if it did, the additional removal of the candidate's neighbor rectified the situation. Belavin et al's model is simple and effective for generating vacancies, but its physical basis as a 2D spatial random field model is not clear and it cannot be extended to areal defects involving more than one atom such as a SW defect.

Since there is not enough information in the experimental literature to provide a clear picture of statistical properties of the defects, it is reasonable to start with the assumption that the defects occur in a completely random manner, which implies an underlying homogeneous Poisson spatial process. We also incorporate the fact that the SW defect is not a point defect but has a finite area and there should be no overlap between neighboring defects. Therefore, we adopt a Matern hard-core point process [62] for the defect field. A Matern hard-core process is simply a thinned Poisson point process in which the constituent points are forbidden to lie closer than a minimum distance $h$.

The intensity (i.e., rate of occurrence) of the Matern hard-core process is $\lambda_h = p_h \lambda$ where $\lambda$ is intensity of the underlying homogeneous Poisson point process and $p_h$ is the probability that an arbitrary point from the underlying Poisson process will survive the Matern thinning. The average number of SW defects on an area $A_t$ is thus $\lambda_h A_t$. For a finite tube of length $b$, the probability $p_h$ can be computed as:

$$p_h = \frac{1}{b}\int_0^b \frac{1-e^{-\lambda A(y)}}{\lambda C(y;h)} dy \qquad (1)$$

where $C$ is the area over which a Poisson point at $(x_0, y_0)$ searches for its neighbors [63].

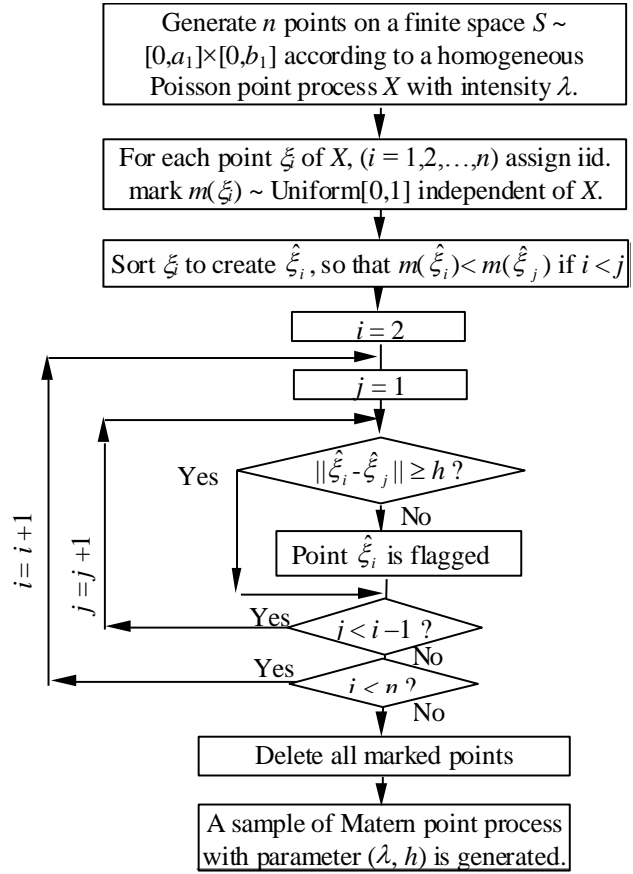

**Figure 1 Generating a Matern hard-core field**

In this study, we fix $h$ at 8.0 Å, and use a set of reasonable values for $\lambda$. The algorithm for generating SW defects is outlined in Figure 1. An application of the algorithm is illustrated on a graphene sheet in the following. In this particular example, the graphene sheet is 49.2 Å long and 25.5 Å wide and is used to create a (6,6) armchair SWNT. The SWNT ends up with six SW defects.

Random points are generated uniformly on the graphene sheet with rate $\lambda$ per unit area

(to conform to a Poisson spatial process) and the points are assigned integer id's in the order they are generated (Figure 2 (a)). 8 Poisson points are generated in this example. The distance between every pair of points is checked. If a pair of points are less than $h$ apart, the point with the smaller id in that pair is marked ((Figure 2 (b)). Once all pairs are compared, the marked points are deleted, thus producing a Matern hard-core field. In this example, two points (numbers 4 and 5) are deleted to produce the hard-core field. For each surviving point, the closest carbon-carbon bond on the graphene sheet is located (Figure 2 (c)) and the bond is rotated by 90° to form an SW defect (Figure 2 (d)). The graphene sheet with the six SW defects are shown in Figure 3(a); the corresponding armchair SWNT is shown in Figure 3(b).

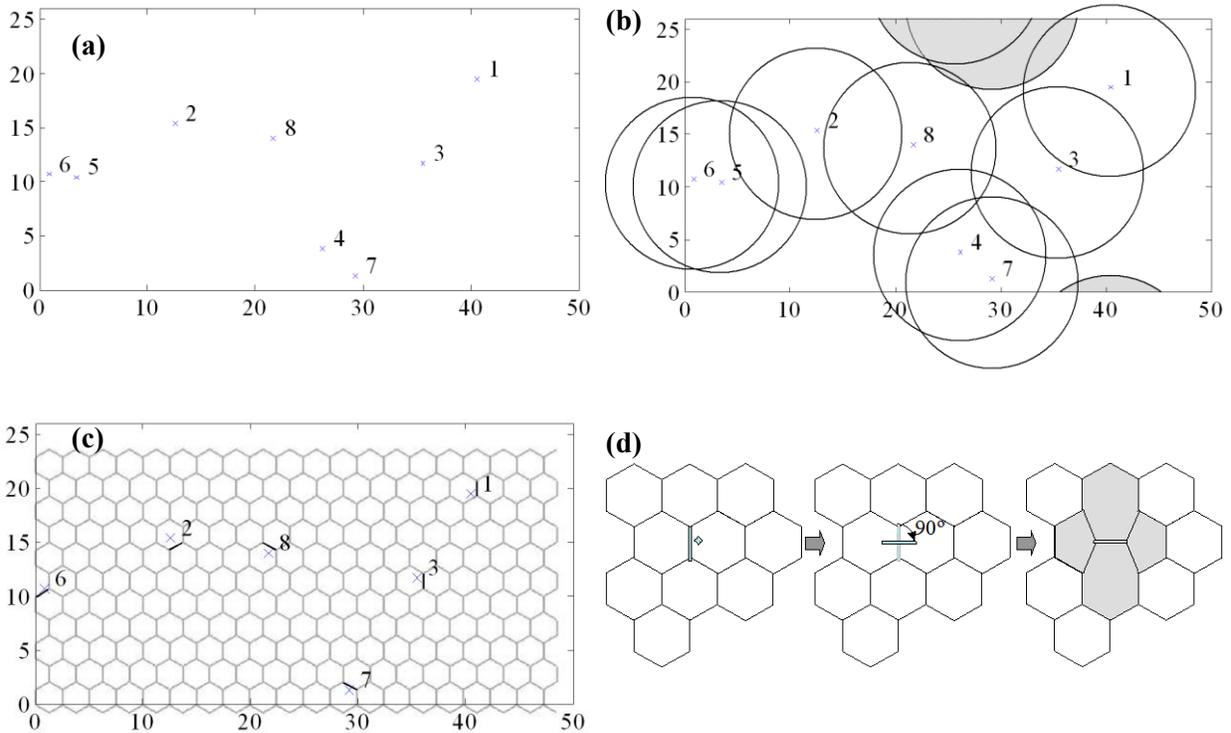

**Figure 2: Generation of Stone-Wale defects on a graphene sheet. (a) Poisson points are generated on a plane. (b) Pairs of points less than h apart are identified. (c) The Matern hard-core field is produced by thinning the Poisson field (d) SW defects are generated by rotating C-C bond**

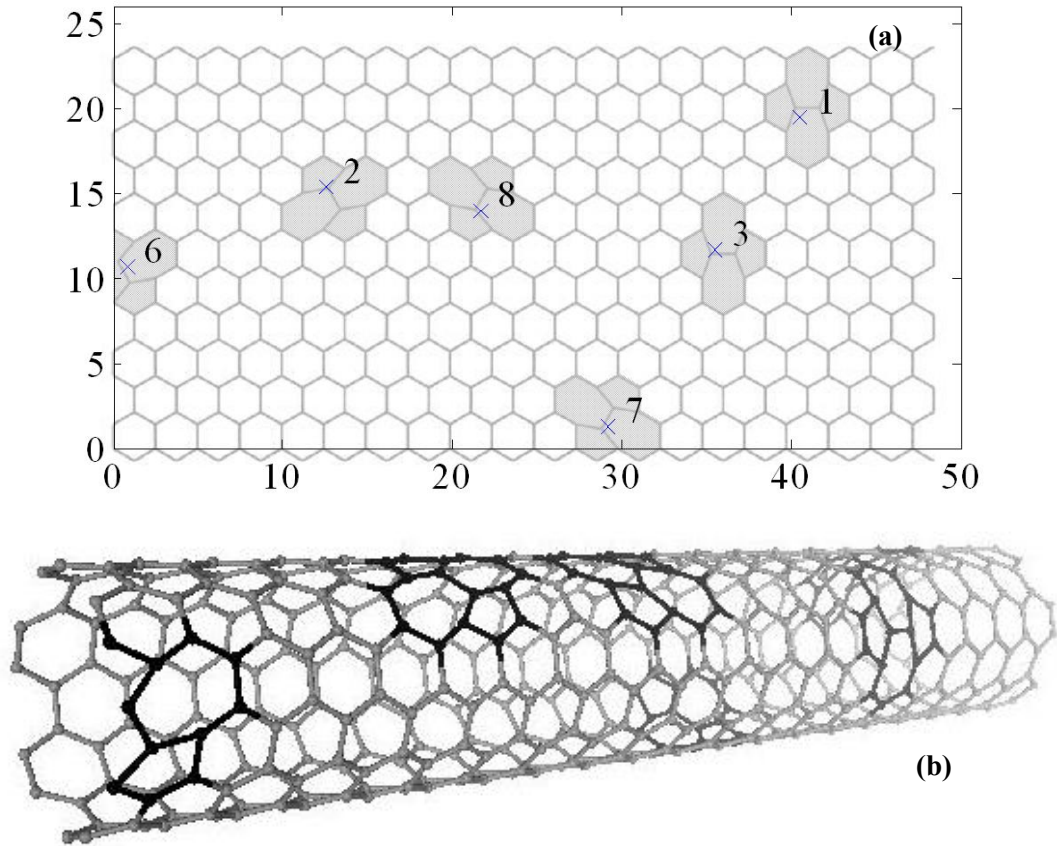

**Figure 3: (a) Graphene sheet with six SW defects, (b) the corresponding (6,6) SWNT**

### 3. Mechanical behavior of SWNTs with random SW defects

Due to the small size of carbon nanotubes, laboratory experiments to measure their mechanical properties are difficult and potentially expensive at the current state of the art. Atomistic simulation or AS, therefore, is a very attractive tool that can complement laboratory experiments in studying the mechanical behavior and failure of small scale structures such as carbon nanotubes, particularly when heterogeneities and the discrete nature of the material becomes dominant [64]. Although AS has a natural advantage over classical continuum based approaches in studying small-scale mechanics, we should also mention that elegant continuum approaches have been used to study the mechanical properties of carbon nanotubes [17, 65-70]. These continuum approaches are computationally efficient compared with atomistic simulation, however, they both have limitations. They apply well to the *defect-free* nanostructure; the quasi-continuum approach is especially preferred when the deformation is uniform, so that a closed form strain energy can be derived and the mechanical properties can be solved analytically. And they are both intrinsically static (the kinetic energy of the atoms is not considered) making it more difficult to study dynamic properties. Finally, to the knowledge of the authors, these continuum based methods have not been applied to study properties of nanotubes in the presence of randomness.

For the atomistic simulation part of this study, a modified Morse potential model for describing the interaction among carbon atoms [71] is applied. This potential model does not have some of the shortcomings of the bond order potential models, for example, the cutoff function in Brenner's bond order model is said to give rise to spurious forces and inaccurately large breaking strain [30, 60, 72].

The bond-breaking criterion is an important issue in the simulation of fracture of solids. Most atomistic simulation studies adopt a distance-based criterion ($r_f$) for this, i.e., the bond between atoms is regarded as broken if the separation between atoms exceeds a critical distance [73-77]. A good guess is that $r_f$ is equivalent to the cut-off distances $r_c$ of potential energy functions. Based on studies such as [31, 59], we adopt the cutoff distance ($r_c$) and the critical inter-atomic separation ($r_f$) as: $r_f = r_c = 1.77$ Å in this example.

In order to study the effects of Stone-Wales defects on mechanical properties and the fracture process of carbon nanotubes, we adopt two single-walled nanotubes (SWNT) configurations in this study: the (6,6) armchair having diameter 8.14Å and the (10,0) zigzag having diameter 7.82Å. The total number of atoms in the simulation was 460 for the 49.2 Å long (10,0) zigzag tube and 480 for the 49.2 Å long (6,6) armchair tube.

The initial atomic positions are obtained by wrapping a graphene sheet (a typical example was shown in Figure 3) into a cylinder along the chiral vector $\mathbf{C}_n = m\mathbf{a}_1 + n\mathbf{a}_2$ such that the origin (0,0) coincides with the point $(m, n)$. The distance between neighboring carbon atoms on the graphene sheet, $a_0$, is 1.42 Å, which is the C-C sp$^2$ bond length in equilibrium. The tube diameter is thus obtained as $d = a_0\sqrt{3(m^2 + n^2 + mn)}/\pi$.

The initial atomic velocities are randomly chosen according to a uniform distribution (between the limits −0.5 and 0.5) and then rescaled to match the initial temperature (300K in this example). We assume that the nanotube is in vacuum or in air, and the heat conduction between the tube and the environment is relatively small compared to the heat accumulation from the mechanical loading at high speeds. From test runs, the temperature fluctuations in the simulations are found to be within acceptable ranges. Therefore, no temperature control is implemented in this study. The mechanical loading is uniaxial and displacement controlled: it is applied by moving the atoms at either end (12 atoms at each end, as shown in Figure 4) in the axial direction away from each other at constant speed (10 m/s) without relaxing until the occurrence of fracture.

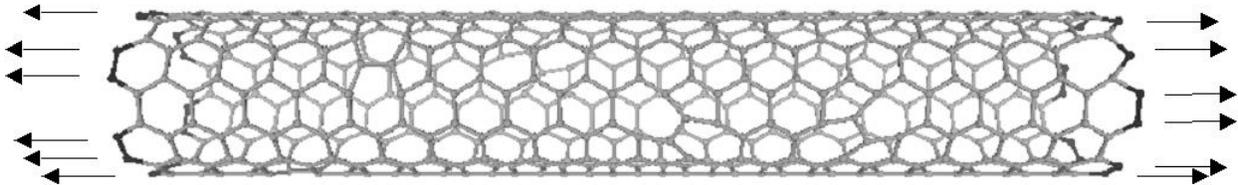

**Figure 4 Mechanical loading is applied through moving the outermost atoms at both ends (highlighted at bond ends)**

Three mechanical properties are calculated from the simulated force and displacement time histories: (i) the Young's modulus is calculated as the initial slope of the force-displacement curve; (ii) the ultimate strength is calculated at the maximum force point, $\sigma_u = F_{max}/A$, where $F$ is the maximum axial force, $A$ is the cross section area, assuming the thickness of tube wall is 0.34 nm; (iii) the ultimate strain, which corresponds to the ultimate strength, is calculated as $\varepsilon_u = \Delta L_u/L$, where $L$ is the original tube length. Other important information can also be obtained through the simulation of tensile test of SWNTs, such as the time histories of energies and bond angles, dependence of mechanical properties on diameter, chirality and loading rate, etc. [17-19, 21, 32, 64].

Examples of the force-displacement relations of a (6,6) SWNT are shown in Figure 5, with the same initial velocity distribution in each case but with different numbers of SW defects (0, 2, 4, 6, respectively). The displacement controlled loading rate is 10 m/s. The reduced units (r.u.) used in this and subsequent figures have the following physical equivalents: 1 time r.u. = 0.03526 pico-second; 1 force r.u. = 1.602 nano-Newton; 1 displacement r.u. = 1 Angstrom.

It is clear from Figure 5 that defects can have significant impact on mechanical behavior of nanotubes: SWNTs with more defects are likely to break at smaller strains and have less strength as well. The force-displacement curve in each case in Figure 5 behaves almost linearly up to about half way, although there is no obvious yield point. The slope then begins to decrease up to the breaking point where an abrupt drop of force occurs. The effect of defects on the Young's modulus, however, appears much less pronounced.

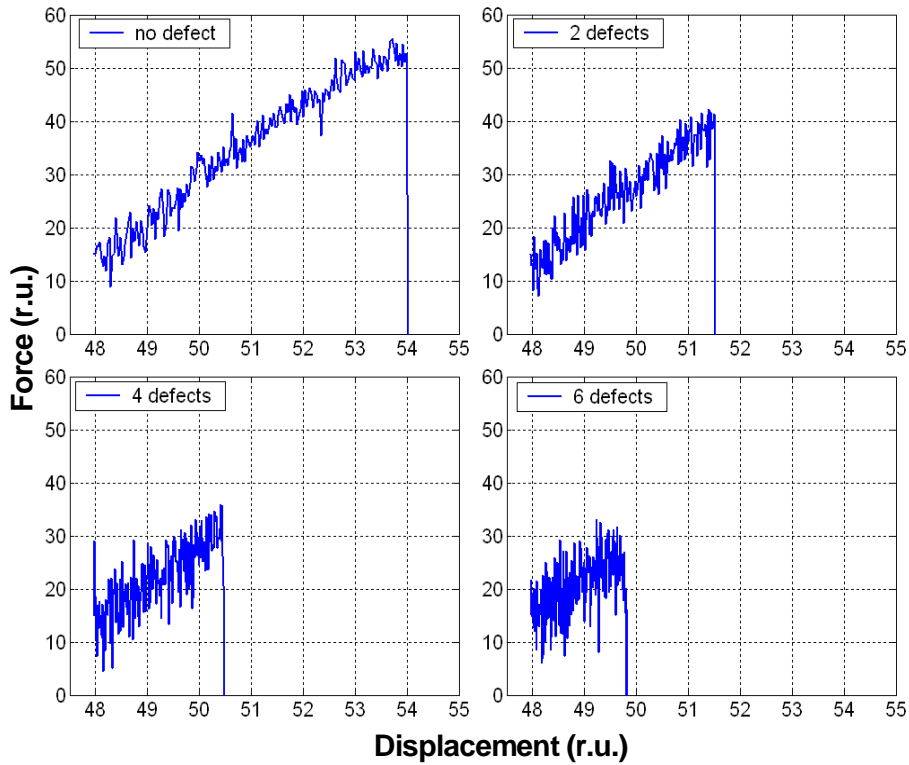

**Figure 5 Force-displacement curves of nanotubes with various average number of defects**

Eight snapshots taken from the fracture process of a defect-free (6,6) armchair SWNT are shown in Figure 6(a), illustrating the detailed crack evolution from the first bond breaking to the complete separation into two parts. Figure 6(b) shows the corresponding time instants of these snapshots on the force time history. The fracture process of a similar (6,6) armchair SWNT but with three Stone-Wales defects are studied next (Figure 7(a)). The corresponding force time history is shown in Figure 7(b), showing that the SWNT breaks much earlier than the defect-free tube under identical loading. The crack initiates from a heptagon, and grows irregularly until it breaks the tube. This is representative of all subsequent tensile simulations of SWNTs with defects: cracks invariantly initiate from a defect and grows irregularly. For a defect-free tube, in contrast, the crack initiates at quite random locations (as shown in Figure 8) on the tube surface and grows at an angle of about 60° with the tube axis in case of (6,6) armchair, and at about 90° with the tube axis in case of (10,0) zigzag. The only source of randomness in Figure 8 is in the initial velocity distribution of the atoms. The tube length in each case in Figure 5 through Figure 8 is constant at 49.2 Å as is the displacement controlled loading rate at 10 m/s.

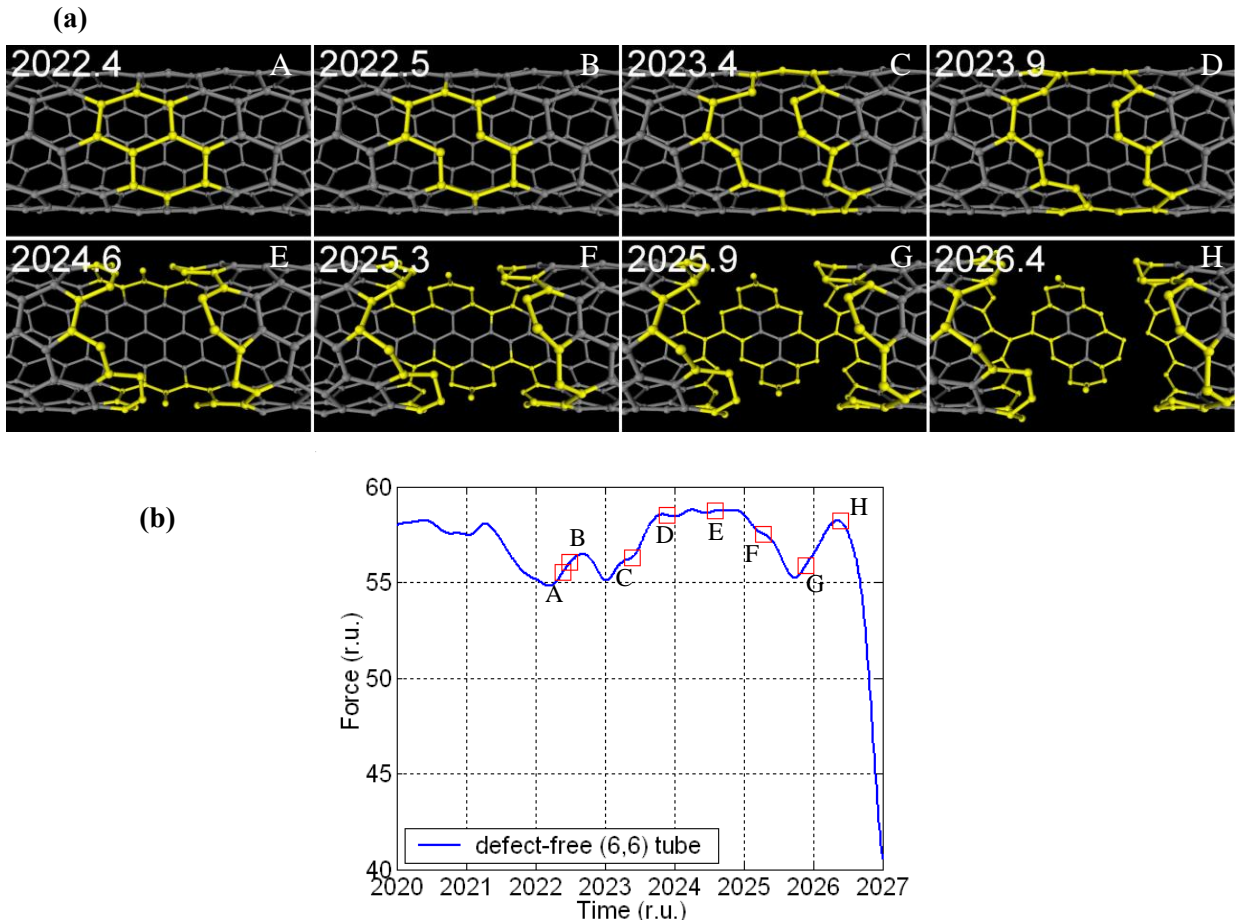

**Figure 6 Fracture process of a defect-free (6,6) SWNT (a) crack initiation and propagation (A–H) (b) corresponding force time history**

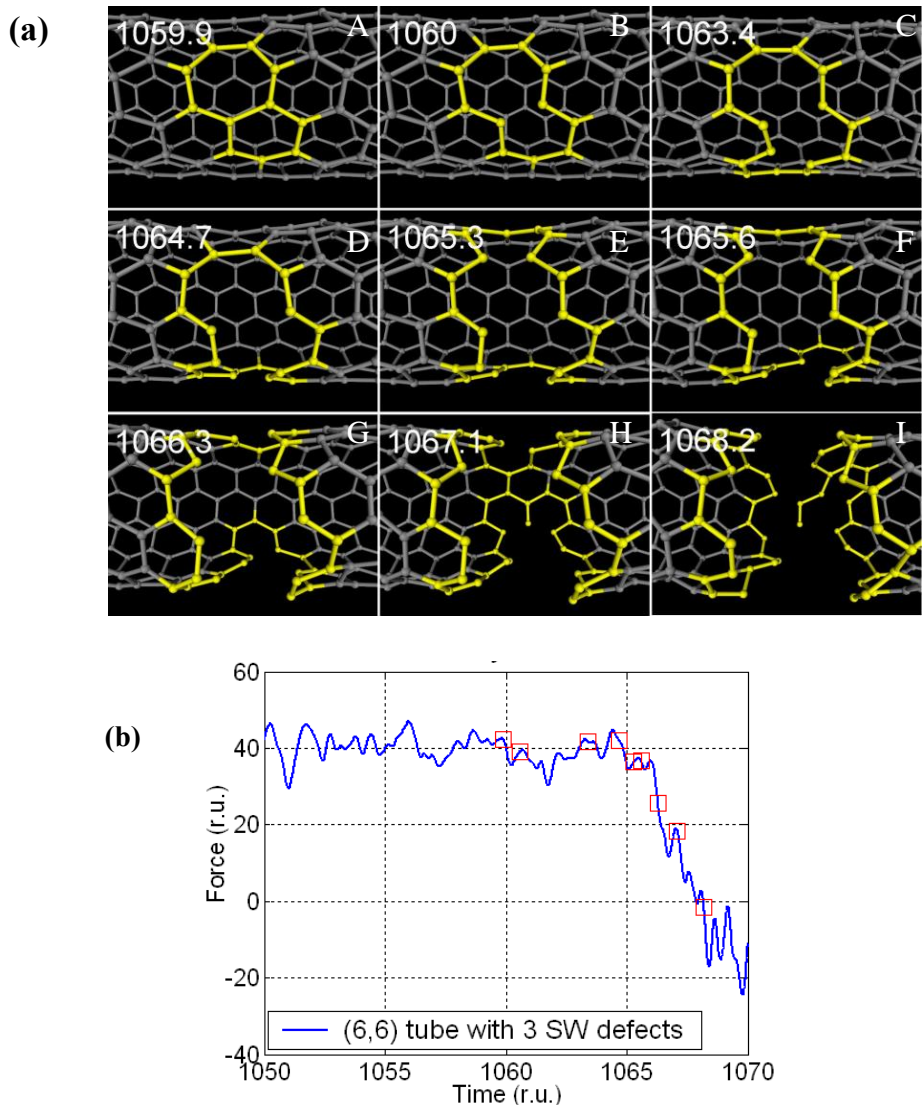

**Figure 7** Fracture process of a (6,6) SWNT with three SW defects (a) crack initiation and propagation (A – I) (b) corresponding force time history

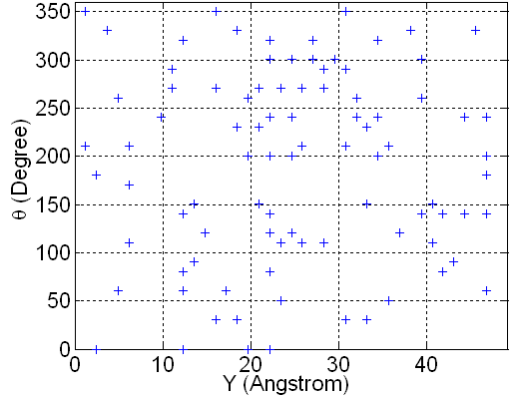

**Figure 8: 2D spatial distribution of crack initiation site in defect-free (6,6) SWNT (99 samples)**

The effects of randomness in the spatial distribution of SW defects on mechanical behavior is now investigated for both the armchair and zigzag SWNTs. The tube length in each case is 49.2 Å while the displacement controlled loading rate is constant at 10 m/s. For each tube, we start with the defect-free case ($\lambda = 0$) and carry out 50 simulations leading to fracture. From these 50 samples we determine the statistics of the SWNT mechanical properties as shown in Table 3. The only source of randomness here is in the distribution of the initial atomic velocities. The resultant c.o.v. of Young's modulus (~1% for armchair and ~2% for zigzag) as well as that of ultimate strength (about 2% for either configuration) are negligible. The c.o.v. of ultimate strain, though higher at around 5%, is still rather small. Hence it is clear that, at the given temperature of 300K, the randomness resulting from initial velocity distribution is insignificant.

We then move on to studying four cases with randomly occurring SW defects corresponding to four different mean field values, starting from an average of 0.9 defects per tube up to 3.9 mean defects per tube. For each mean value, 50 SWNT samples with defects are generated according to a Matern hardcore process, and loading leading to fracture is simulated as before. In order to systematically study the effects of spatial randomness, the initial velocity distributions are kept identical to those in the respective 50 defect-free cases. Figure 9 elaborates the effect of mean number of defects on the SWNT Young's modulus. It is clear that the scatter in the Young's modulus increases with the more defects and tube becomes more compliant as well. The variation in the Young's modulus in Figure 9 agrees with experimental results reported in Table 1and Table 2. Complete statistics of the mechanical properties of the tubes with defects are listed in Table 3 and the trend is graphed in Figure 10.

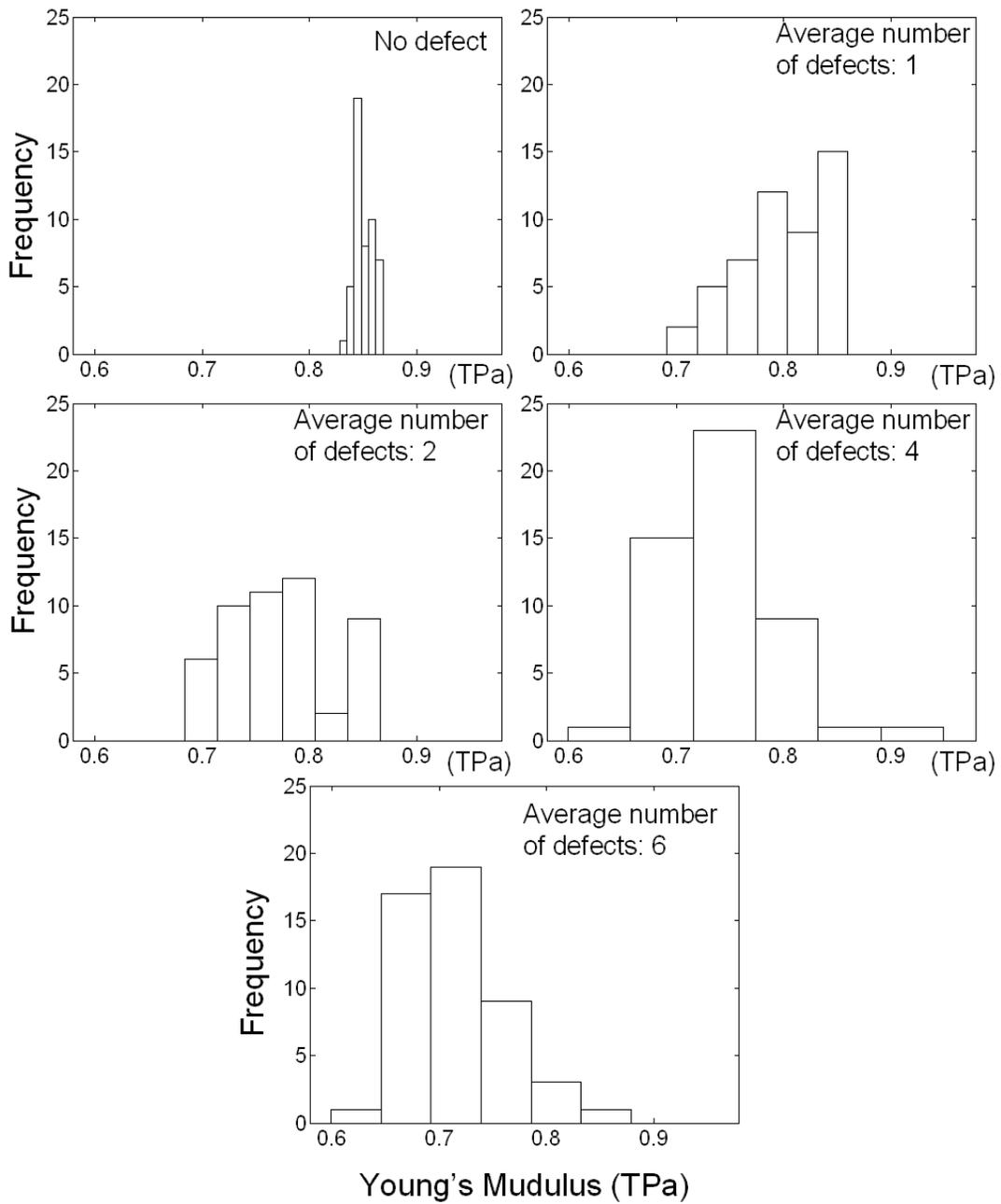

**Figure 9: Distribution of Young's modulus of (6,6) armchair nanotubes as a function of average number of SW defects (50 samples each)**

**Table 3 Young's modulus, ultimate strength and ultimate strain of SWNTs with different SW defect intensities**

| Mean Poisson points $\lambda A_t$ | Mean Matern disks $\lambda_h A_t$ | $E^*$ (TPa) | | $\sigma_u^*$ (GPa) | | $\varepsilon_u^*$ | |
|---|---|---|---|---|---|---|---|
| | | μ | V | μ | V | μ | V |
| (6,6) Armchair | | | | | | | |
| 0 | 0 | 0.851 | 1.01% | 105.5 | 2.01% | 12.45% | 4.93% |
| 1.0 | 0.925 | 0.793 | 5.50% | 89.71 | 13.80% | 8.69% | 33.91% |
| 2.0 | 1.716 | 0.771 | 6.25% | 83.11 | 13.52% | 7.20% | 36.05% |
| 4.0 | 2.970 | 0.738 | 7.77% | 76.29 | 10.02% | 5.50% | 29.56% |
| 6.0 | 3.889 | 0.716 | 6.76% | 74.56 | 8.93% | 4.94% | 31.61% |
| (10,0) Zigzag | | | | | | | |
| 0 | 0 | 0.831 | 2.17% | 86.78 | 1.85% | 8.88% | 4.89% |
| 1.0 | 0.925 | 0.778 | 7.15% | 74.53 | 12.52% | 5.71% | 39.85% |
| 2.0 | 1.716 | 0.775 | 7.89% | 70.17 | 14.75% | 4.75% | 49.56% |
| 4.0 | 2.970 | 0.750 | 10.17% | 65.80 | 7.12% | 3.57% | 30.25% |
| 6.0 | 3.889 | 0.711 | 14.78% | 63.05 | 9.12% | 2.89% | 43.59% |

*based on statistics of 50 samples; μ = mean; V = coefficient of variation (c.o.v.), $E$ = Young's modulus, $\sigma_u$ = ultimate strength, $\varepsilon_u$ = ultimate strain (strain at max force)

We clearly notice that the mean values of stiffness, strength and ultimate strain of the tube for either configuration decrease as the average number of defects increases. However, the uncertainty associated with each property (quantified by the c.o.v.) does not show such monotonic trend in each case. Except for the Young's modulus of the zigzag tube, it is rather interesting to note that the c.o.v.'s of these properties reach their respective maxima when the average number defects is in the range 1 ~ 3. Nevertheless, the most striking observation is perhaps that the introduction of an additional defect has the most pronounced effect on the randomness in mechanical properties when the tube is originally defect free than when one or more defects are already present.

The zigzag tube differs from the armchair one in two ways. For a given mean number of defects per tube, (i) the zigzag configuration has less strength and less ductility on the average than the armchair tube although the mean stiffness is about the same; and (ii) the zigzag configuration shows more uncertainty in its stiffness and ultimate strain compared with the armchair tube although the strength has roughly the same c.o.v. in either configuration. The fact that zigzag tube has less ultimate strain has been reported by a few studies [46, 56] and the widely observed fact that the brittle fracture mode is more sensitive to random defects has been corroborated by the above findings.

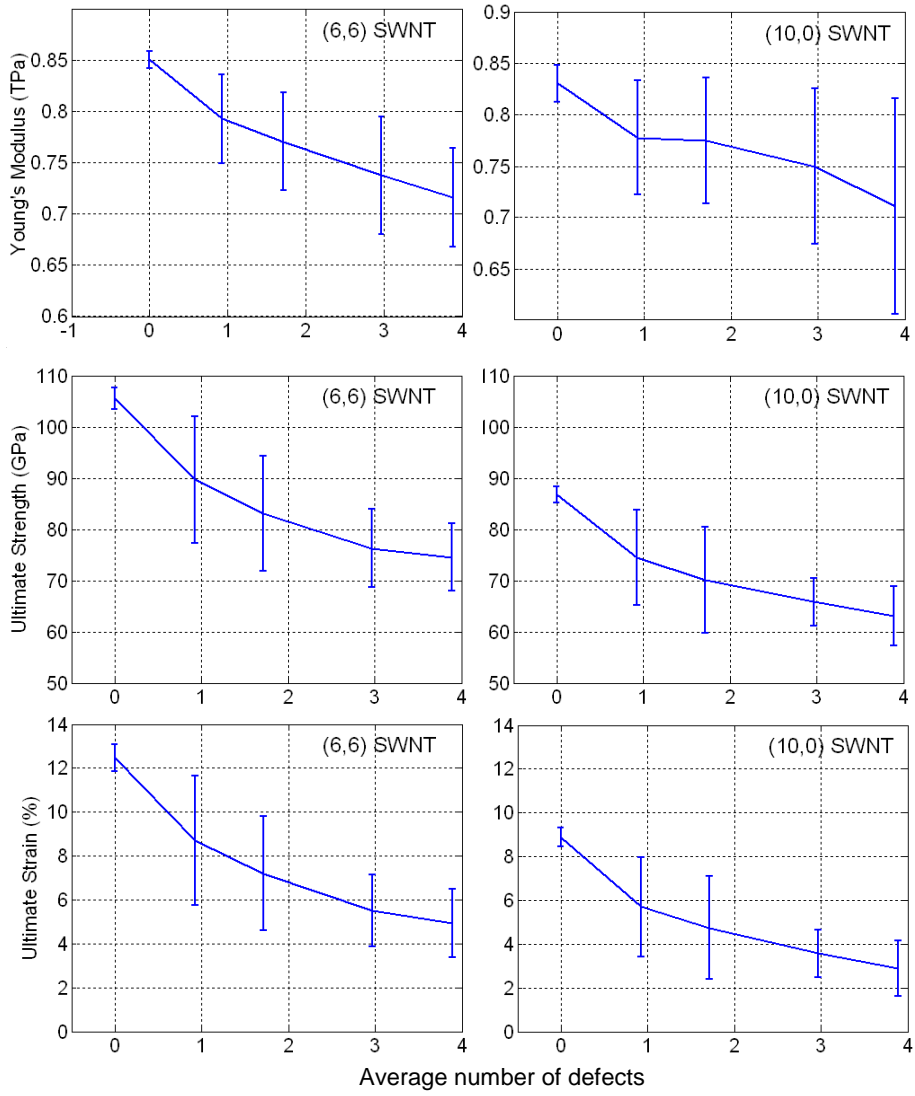

**Figure 10 Effects of random SW defects on mechanical properties of armchair and zigzag SWNTs ($l$ = 49.2 Å, 50 samples each). Solid line represents mean value, vertical bar implies +/- one standard deviation.**

No dislocations similar to what reported in other simulations [46, 47] was observed in our simulations. This may be due to the difference in the potential models and/or the loading modes: our tensile loading is continuous and monotonic at room temperature and we used a modified Morse potential, while Nardelli et al. [47] used a Bond order potential model and allowed unloading and relaxation at high temperature.

## 4. Summary and Conclusions

Carbon nanotubes are known to have ultra-high stiffness and strength, yet a wide scatter has been observed in the mechanical properties of carbon nanotubes. It is likely that defects, which are commonly present in CNTs, may have significant effects on the mechanical and other properties of CNTs. In this paper we considered the random nature of Stone-Wales (SW or 5-7-7-5) defects in CNTs and, through the technique of atomistic simulation, demonstrated that randomly occurring defects can have significant influence on the fracture process and mechanical properties of SWNTs. A Matern hard-core random field applied on a finite cylindrical surface was used to describe the spatial distribution of the Stone-Wales defects.

We simulated a set of displacement controlled tensile loading up to fracture of SWNTs with (6,6) armchair and (10, 0) zigzag configurations. The force-displacement curve typically behaved almost linearly up to about half way, although there was no obvious yield point. The stiffness, strength and ultimate strain were calculated from the simulated force and displacement time histories. We found that fracture invariably initiated from a defect if one was present and the crack grew irregularly. For a defect-free tube, the only source of randomness was in the initial velocity distribution of the atoms, and the crack initiated at quite random locations on the tube surface.

The effects of randomness in the spatial distribution of SW defects on mechanical behavior was investigated. Along with the defect free-case, we studied four cases with randomly occurring SW defects starting from an average of 0.9 defects per tube up to 3.9 mean defects per tube. The randomness resulting solely from the initial velocity distribution (i.e., the defect free case) was rather insignificant at the operating temperature of 300K. The mean values of stiffness, strength and ultimate strain of the tube decreased as the average number of defects increased. However, the uncertainty associated with each property did not show such monotonic trend in each case. The introduction of an additional defect had the most pronounced effect when the tube was originally defect free than when one or more defects were already present. The zigzag tube was found to more brittle and to exhibit more uncertainty in its mechanical properties than the armchair one.

Future work should be extended to more types of defects such as vacancies, heterogeneous atoms, etc. and more complicated structures. Experimental studies on statistical distribution of such defects and of mechanical properties of nanotubes are also required for validation purposes.

## 5. Acknowledgment

A set of preliminary results from this work was presented at the 17$^{th}$ ASCE Engineering Mechanics Conference held at the University of Delaware, Newark, DE, USA in June 2004.## 6. References other properties of CNTs. In this paper we considered the random nature of Stone-Wales (SW or 5-7-7-5) defects in CNTs and, through the technique of atomistic simulation, demonstrated that randomly occurring defects can have significant influence on the fracture process and mechanical properties of SWNTs. A Matern hard-core random field applied on a finite cylindrical surface was used to describe the spatial distribution of the Stone-Wales defects.

We simulated a set of displacement controlled tensile loading up to fracture of SWNTs with (6,6) armchair and (10, 0) zigzag configurations. The force-displacement curve typically behaved almost linearly up to about half way, although there was no obvious yield point. The stiffness, strength and ultimate strain were calculated from the simulated force and displacement time histories. We found that fracture invariably initiated from a defect if one was present and the crack grew irregularly. For a defect-free tube, the only source of randomness was in the initial velocity distribution of the atoms, and the crack initiated at quite random locations on the tube surface.

The effects of randomness in the spatial distribution of SW defects on mechanical behavior was investigated. Along with the defect free-case, we studied four cases with randomly occurring SW defects starting from an average of 0.9 defects per tube up to 3.9 mean defects per tube. The randomness resulting solely from the initial velocity distribution (i.e., the defect free case) was rather insignificant at the operating temperature of 300K. The mean values of stiffness, strength and ultimate strain of the tube decreased as the average number of defects increased. However, the uncertainty associated with each property did not show such monotonic trend in each case. The introduction of an additional defect had the most pronounced effect when the tube was originally defect free than when one or more defects were already present. The zigzag tube was found to more brittle and to exhibit more uncertainty in its mechanical properties than the armchair one.

Future work should be extended to more types of defects such as vacancies, heterogeneous atoms, etc. and more complicated structures. Experimental studies on statistical distribution of such defects and of mechanical properties of nanotubes are also required for validation purposes.

## 5. Acknowledgment

## 6. References

[1]. Broek, D., *Elementary engineering fracture mechanics*. 4th ed. 1991: Kluwer Academic.
[2]. Ebbesen, T.W. and P.M. Ajayan, *Large-scale synthesis of carbon nanotubes.* Nature, 1992. **358**: p. 220.
[3]. Hamada, N., S. Sawada, and A. Oshiyama, *New one-dimensional conductors: Graphitic microtubules.* Physical Review Letters, 1992. **68**(1579-1582).
[4]. Saito, R., et al., *Electronic structure of graphene tubules based on C60.* Physical Review B, 1992. **46**: p. 1804–1811.
[5]. Saito, R., et al., *Electronic structure of chiral graphene tubules.* Applied Physics Letters, 1992. **60**(2204).
[6]. Treacy, M.M., et al., *Exceptional high Young's modulus observed for individual carbon nanotubes.* Nature, 1996. **381**(6584): p. 678.
[7]. Wong, E.W., P.E. Sheehan, and C.M. Lieber, *Nanobeam mechanics: Elasticity, strength, and toughness of nanorods and nanotubes.* Science, 1997. **277**(5334): p. 1971-1975.
[8]. Falvo, M.R., et al., *Bending and buckling of carbon nanotubes under large strain.* Nature,

**Figure captions**

Figure 1 Generating a Matern hard-core field

Figure 2: Generation of Stone-Wale defects on a graphene sheet. (a) Poisson points are generated on a plane. (b) Pairs of points less than h apart are identified. (c) The Matern hard-core field is produced by thinning the Poisson field (d) SW defects are generated by rotating C-C bond

Figure 3: (a) Graphene sheet with six SW defects, (b) the corresponding (6,6) SWNT

Figure 4 Mechanical loading is applied through moving the outermost atoms at both ends (highlighted at bond ends)

Figure 5 Force-displacement curves of nanotubes with various average number of defects

Figure 6 Fracture process of a defect-free (6,6) SWNT (a) crack initiation and propagation (A−H) (b) corresponding force time history

Figure 7 Fracture process of a (6,6) SWNT with three SW defects (a) crack initiation and propagation (A – I) (b) corresponding force time history

Figure 8: 2D spatial distribution of crack initiation site in defect-free (6,6) SWNT (99 samples)

Figure 9: Distribution of Young's modulus of (6,6) armchair nanotubes as a function of average number of SW defects (50 samples each)

Figure 10 Effects of random SW defects on mechanical properties of armchair and zigzag SWNTs ($l$ = 49.2 Å, 50 samples each). Solid line represents mean value, vertical bar implies +/- one standard deviation.